\documentclass[twocolumn,superscriptaddress,nofootinbib,floatfix,aps,showpacs,prb,citeautoscript,reprint]{revtex4-1}
\usepackage{dcolumn}
\usepackage{bm}
\usepackage{amsmath}
\usepackage{amsfonts}
\usepackage{amssymb}
\usepackage{graphicx}
\usepackage{tabularx}
\usepackage{subfigure}
\usepackage{color}
\usepackage{gensymb}
\usepackage{physics}
\usepackage{multirow}
\usepackage{booktabs}
\usepackage[utf8]{inputenc}
\usepackage{threeparttable}
\usepackage{tablefootnote}
\usepackage{adjustbox}
\usepackage[symbol]{footmisc}

\usepackage{hyperref} \hypersetup{colorlinks=true, linkcolor=blue,
  citecolor=blue, urlcolor=blue, pdftitle={The elastic, mechanical and thermodynamic properties of Bi-Sb binaries}}

\begin{document}
\title{Effect of spin-orbit coupling on the elastic, mechanical, and thermodynamic properties of Bi-Sb binaries}

\author{Sobhit Singh } 
\email{smsingh@mix.wvu.edu}
\affiliation{Department of Physics and Astronomy, West Virginia University, Morgantown, WV-26505-6315, USA}

\author{Irais Valencia-Jaime}
\affiliation{Department of Chemistry, University of North Dakota, Grand Forks, ND 58202, United States}

\author{Olivia Pavlic}
\affiliation{Department of Physics and Astronomy, West Virginia University, Morgantown, WV-26505-6315, USA}

\author{Aldo H. Romero}
\affiliation{Department of Physics and Astronomy, West Virginia University, Morgantown, WV-26505-6315, USA}

\begin{abstract}
Using first principles calculations, we systematically study the elastic stiffness constants, mechanical properties, elastic wave velocities, Debye temperature, melting temperature, and specific heat of several thermodynamically stable crystal structures of Bi$_{x}$Sb$_{1-x}$  ($0 < x < 1$) binaries, which are of great interest due to their numerous inherent rich properties, such as thermoelectricity, thermomagnetic cooling, strong spin-orbit coupling (SOC) effects, and topological features in the electronic bandstructure. We analyze the bulk modulus ($B$), Young's modulus ($E$), shear modulus ($G$), $B/G$ ratio, and Poisson's ratio ($\nu$) as a function of the Bi concentration in Bi$_{x}$Sb$_{1-x}$. The effect of SOC on above mentioned properties is further investigated. In general, we observe that the SOC effects cause elastic softening in most of the studied structures. Three monoclinic structures of Bi-Sb binaries are found to exhibit significantly large auxeticity. The Debye temperature and the magnitude of the elastic wave velocities monotonically decrease with increasing Bi-concentration. We also discuss the specific heat capacity versus temperature data for all studied binaries. Our theoretical results are in excellent agreement with the existing experimental and theoretical data. The comprehensive understanding of the material properties such as hardness, mechanical strength, melting temperature, propagation of the elastic waves, auxeticity, and heat capacity is vital for practical applications of the studied binaries. 
\end{abstract}

\date{\today}


\maketitle
\section{Introduction}

A thorough understanding of the mechanical response of any given material is essential before the technological applications of that particular material can be realized. A good place to start is investigating the elasticity, a fundamental property of a crystal which governs the macroscopic response of the crystal under external forces. The hardness, mechanical strength and the propagation of the sound and elastic waves in a given material can be determined by knowledge of the elastic constants of that particular material. Amongst many known binary compounds and alloys, Bi-Sb based binaries have retained a peculiar place due to their applications in the low-temperature thermoelectric industry and refrigeration. \cite{BiSb-thermo1985, lenoir1996bi, YIM1972, LENOIR1996_BiSb, Lv2010} Moreover, Bi-Sb binaries are the first predicted three dimensional topological insulator (often referred as the first generation topological insulator) that host robust conducting surface states. \cite{FuPRB2007, Hseih2008, TeoPRB2008, Zhong2009, KaneRev2010} Soon after the theoretical prediction, Hseih et al. \cite{Hseih2008} reported the experimental detection of novel gapless conducting surface states in this binary system. Recently, we found that the lowest energy structure of BiSb composition (in $R3m$ space group) exhibits large ferroelectric behavior along with a giant tunable Rashba-Dresselhaus effect, which is the result of the broken inversion symmetry and the large spin-orbit coupling (SOC) of the constituent Bi and Sb elements. \cite{sobhit2016PRB, singh2016PCCP} Furthermore, we demonstrated that one can realize a novel Weyl semimetallic phase under external stress of 4--6 GPa.  \cite{sobhit2016PRB} Interestingly, by exploiting an interlink between the large SOC of the constituent atoms and the ferroelectric polarization, one can tune the dynamics of Weyl fermions in the momentum space of BiSb. This particular property is of notable interest for applications of Weyl semimetals in the forthcoming Weyltronic technology. 

BiSb is not only interesting in its bulk phase, but it also shows unique electronic properties in two-dimensions. \cite{Tan2017, Yu2017, MBrzezi2017, sobhitPRB_2DBiSb_monolayer} In particular, a giant tunable Rashba effect along with a large direct bandgap ($\sim$1.6 eV) has been reported for this system. \cite{sobhitPRB_2DBiSb_monolayer} The existence of the large tunable Rashba effect together with a direct bandgap in the visible region makes this material of peculiar interest for its applications in the optoelectronics and spintronics industry. Recently, Yu et al. \cite{Yu2017} investigated the topological properties of monolayer BiSb, and observed the emergence of robust novel quantum spin Hall (QSH) effect under biaxial tensile strain. This finding was further confirmed by the calculation of Z2 topological invariant and the nontrivial topological edge states. These features make BiSb an attractive candidate for applications in spintronic devices. 

Other than the BiSb composition, several other stable compositions of the Bi-Sb binaries have been reported in literature by both theoretical and experimental studies. \cite{LeePRB2014, singh2016PCCP, Gonzalo2006, dismukes1968lattice, Berger1982,  lenoir1996bi, YIM1972, BiSb-thermo1985, Lv2010} The formation mechanism and the chemical synthesis procedure of Bi$_{x}$Sb$_{1-x}$ nanocrystals are given in refs.\cite{Datta2011, Zhang2013, Kaspar2017}  A detailed structural, electronic, vibrational and thermoelectric investigation of the Bi-Sb binaries can be found in our recent work. \cite{singh2016PCCP} In ref.\cite{singh2016PCCP}, we explored the potential energy surface of Bi-Sb binaries using the minima hopping method \cite{Goedecker2004jpc, Amsler2010jpc}, and calculated the theoretical convex-hull of Bi--Sb. We not only discovered several new energetically and thermodynamically stable crystal structures of Bi-Sb binaries that are located on the convex-hull, but we also recovered the known structures of Bi-Sb binaries in our structural search calculations. In the present work, we investigate the elastic and thermodynamic response of the stable Bi-Sb binaries. All the studied structures could be synthesized in laboratory under suitable ambient conditions. \cite{Lv2010, singh2016PCCP, Datta2011, Zhang2013, Kaspar2017} 

Changes in the mechanical properties of Bi$_{x}$Sb$_{1-x}$ single crystals as a function of Sb-concentration have been studied by ultrasonic wave velocities measurements at room temperature as well as at low-temperatures.\cite{Gopinathan1974, Lichnowski1976, Cankurtaran1985, Dominec1985, Cankurtaran1986, EMUNA2013} In general, the elastic properties, {\it i.e.} bulk modulus, Young's modulus, and shear modulus increase with increasing Sb-concentration in Bi-Sb binaries. \cite{Gopinathan1974} Also, the average speed of sound increases with increasing Sb-concentration, however, it decreases with increasing temperature. \cite{EMUNA2013} Although, most of the experiments report monotonous increase in the elastic moduli with increasing Sb-concentration for larger atomic $\%$ of Sb, there exists some anomalies in the variation of the mechanical properties at low Sb concentration, which is consistent with our theoretical findings. \cite{Gopinathan1974} Although, the specific heat of pristine Bi\cite{KeesomBi1054, CetasPRB1969, Archer1995, BiSOC_PRL2007} and pristine Sb\cite{DeSorbo1953, SbSOC_PRB2008} has been studied in detail,\cite{SbSOC_PRB2008, BiSOC_PRL2007} little attention has been paid to the thermodynamic properties of Bi-Sb binaries\cite{Bunton1969, Lichnowski1976, LeePRB2014}. Lichnowski and Saunders reported increase in Debye temperature with increasing Sb-concentration.\cite{Lichnowski1976} The effect of SOC on the elastic and mechanical properties of Bi-Sb binaries has not yet been reported in the literature, even though SOC is known to significantly change the electronic, vibrational and thermodynamic properties of Bi and Sb based compounds. \cite{BiSOC_PRB2007, SbSOC_PRB2008, BiSOC_PRL2007} 

\begin{figure}[htb!]
 \centering
 \includegraphics[width=8.7cm, keepaspectratio=true]{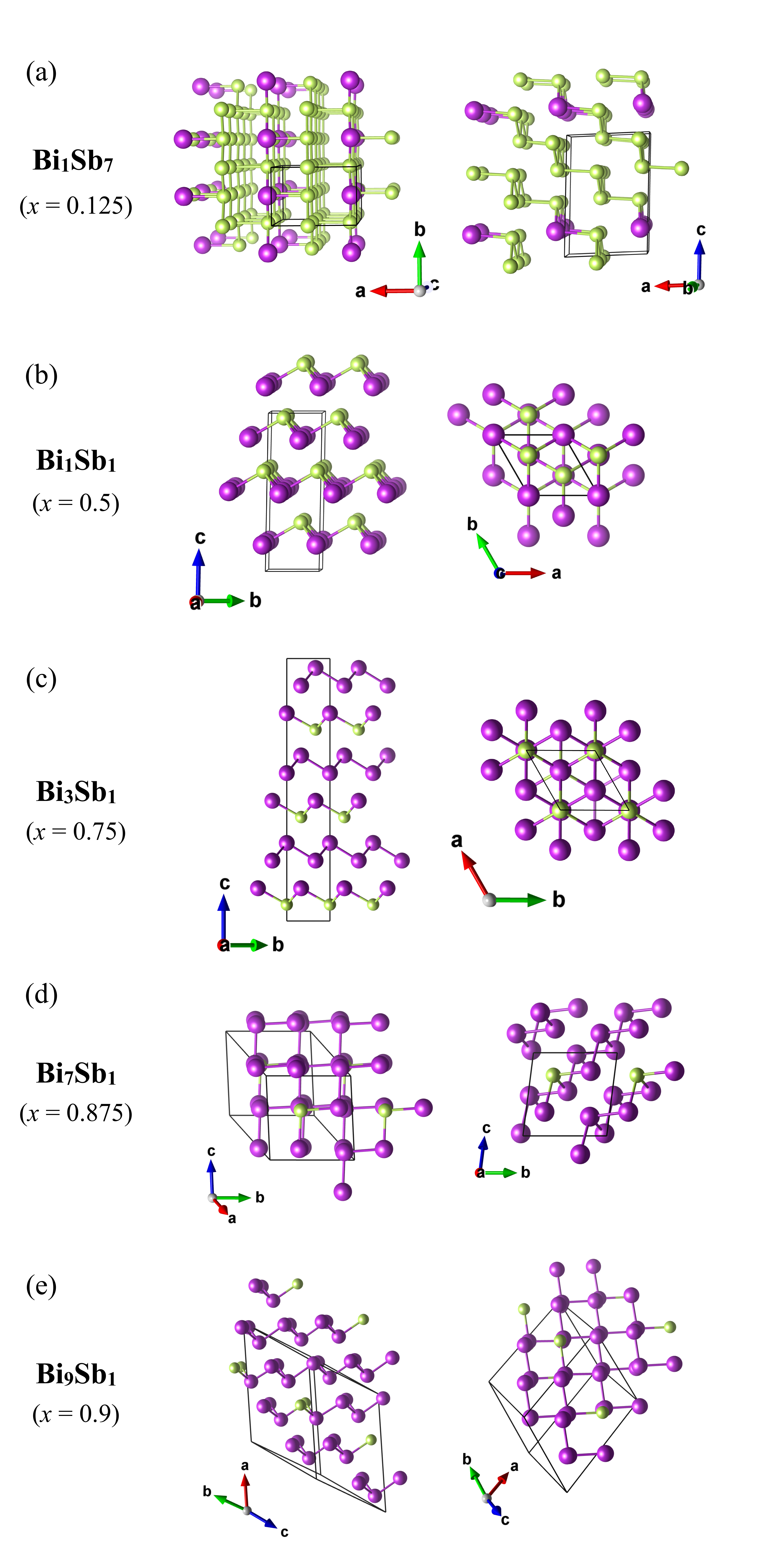}
 \caption{(Color online) Figures (a-e) represent the crystal structure of Bi-Sb binaries located on the Bi--Sb phase diagram\cite{singh2016PCCP}. Bi atoms are shown in purple color while Sb atoms are shown in green color. Each crystal structure is shown from two different lattice orientations. The cutoff length for bonds was defined as 3.10 \AA~in Sb-rich compositions and 3.20 \AA~in Bi-rich compositions. 
 \label{fig:crystal}}
 \end{figure}

In the present work, we report a systematic investigation of the elastic and thermodynamic properties of the Bi-Sb binaries calculated using first-principles. We study the properties of the following Bi-Sb binary compositions (crystal structures are shown in Fig.~\ref{fig:crystal}): Bi$_{1}$Sb$_{7}$, Bi$_{1}$Sb$_{1}$, Bi$_{3}$Sb$_{1}$, Bi$_{7}$Sb$_{1}$ and Bi$_{9}$Sb$_{1}$, which lie on the convex-hull of the Bi-Sb binary phase diagram (the only exception is the Bi$_{3}$Sb$_{1}$ composition which lies above, yet very close to the convex-hull)\cite{singh2016PCCP}. Our results indicate that the ductility of the structures increases with increasing Bi-concentration, whereas in general, the elastic moduli decrease with increasing Bi-concentration. The bulk modulus ($B$), shear modulus ($G$), Young's modulus ($E$), Poisson's ratio ($\nu$) and the elastic stiffness coefficients ($C_{ij}$) of the studied systems are reported below. We notice that Bi$_{1}$Sb$_{7}$, Bi$_{7}$Sb$_{1}$, and Bi$_{9}$Sb$_{1}$ monoclinic structures exhibit negative Poisson's ratio along different spatial directions. The Debye temperature and maximum heat capacity are found to increase with decreasing Bi-concentration. Comparison of our theoretical findings with the available experimental data shows excellent agreement between theory and experiments.

\section{computational details}
Density Functional Theory (DFT) based first-principle calculations were carried out using the projector augmented-wave (PAW) method as implemented in the {\sc VASP} code \cite{Kresse1996, Kresse1999}. We used the PBE exchange-correlation functional as parametrized by Perdew-Burke-Ernzerhof \cite{Perdew1996}. We considered fifteen valence electrons of Bi ($5d^{10}6s^{2}6p^{3}$) and five valence electrons of Sb ($5s^{2}5p^{3}$) in the PAW pseudo-potential. The lattice parameters of each structure were optimized until the Hellmann-Feynman residual forces were less than $10^{-4}$ eV/{\AA} per atom. For convergence of the electronic self-consistent calculations, a total energy difference criterion was defined as $10^{-8}$ eV. We used $650$ eV as kinetic energy cutoff of the plane wave basis set. We employed a $\Gamma$-type $k$-mesh for hexagonal and trigonal structures, while a Monkhorst-pack type $k$-mesh was used to sample the irreducible Brillouin zone of all other crystal phases. The size of $k$-mesh was large enough to ensure the numerical convergence of total energy to less than 1 meV/atom. 

The elastic constants $C_{ij}$ were calculated using the stress-strain relationship as implemented in the {\sc VASP} code. Elastic constants were converged better than 1 GPa by increasing the $k$-mesh size. The bulk modulus ($B$), shear modulus ($G$), Young's modulus ($E$) and Poisson's ratio ($\nu$) quantities were first determined using the Voigt bound\cite{voigt1928} and Reuss bound\cite{reuss1929} schemes, and then an arithmetic average was computed following the Voigt-Reuss-Hill averaging scheme. \cite{Hill1952} This way of evaluating elastic moduli is important since the Voigt and Reuss bounds give an upper and lower estimates of the actual elastic moduli of polycrystalline crystals, respectively. The Voigt bound scheme\cite{voigt1928} relies on the assumption of uniform strain throughout the crystal, whereas the Reuss bound scheme\cite{reuss1929} relies on the assumption of uniform stress throughout the crystal. Since SOC plays an important role in describing the electronic and vibrational properties of Bi and Sb atoms,\cite{BiSOC_PRB2007, SbSOC_PRB2008, BiSOC_PRL2007} we decide to investigate the effect of SOC on the elastic and mechanical properties of Bi-Sb binaries.  Therefore, we have calculated elastic constants and elastic moduli for each studied structure twice: once with-SOC and once without-SOC. The phonopy code\cite{Togo2008, phonopy} was used to calculate the heat capacity of crystal lattice. 

In order to facilitate the analysis of elastic and mechanical properties, we have developed an open source python code named MechElastic\cite{MechElastic}, which can be used to evaluate many important physical quantities such as elastic moduli, elastic wave velocities, Debye temperature, melting temperature, anisotropy factors, and perform the mechanical stability test for any crystalline bulk materials. In future, this code will be generalized for 3D as well as 2D systems.

\section{Results and Discussions}

 \subsection{Elastic constants}

The crystal structures of all the binary compounds under investigation are shown in Fig.~\ref{fig:crystal}. It is important to first discuss the elastic stiffness constants and define their relationship with the macroscopically measurable quantities that give us information about the elastic and mechanical properties of the system. The bulk modulus ($B$), Young's modulus ($E$), shear modulus ($G$), and Poisson's ratio ($\nu$) are known as the elastic moduli and are macroscopically measurable quantities that give a measure of the elasticity of the material. These quantities can be determined from the elastic constants, $C_{ijkl}$. These constants are obtained through the use of the generalized stress-strain Hooke's law\cite{nye1985physical},
\begin{equation}
\sigma_{ij}=C_{ijkl}\epsilon_{kl},
\end{equation}
where $\sigma_{ij}$ and $\epsilon_{kl}$ are the tensile stress and longitudinal strain, respectively. Utilizing the crystal symmetry operations, the total number of constants can be reduced from 81 to 3, 5, 6 and 13 for cubic, hexagonal, tetragonal, and monoclinic structures, respectively. \cite{nye1985physical} Table 1 lists the values of the relevant elastic constants calculated with and without inclusion of SOC. Although, few non-diagonal elements of the $C_{ij}$ matrix contain negative values for Bi$_{7}$Sb$_{1}$ and Bi$_{9}$Sb$_{1}$ binaries, all the six eigen values of the $C_{ij}$ matrix are positive suggesting the elastic stability of these binaries. In fact, all the eigen values of $C_{ij}$ matrix are positive for each studied Bi-Sb binary. 

We notice a small yet significant change in the $C_{ij}$ values due to the SOC effects. Notably, SOC is known to considerably change the electronic and vibrational spectra of Bi and Sb based compounds. Ramifications of SOC on the electronic bandstructure chiefly depend upon the crystal symmetry, and therefore, SOC could have different implications on the same composition but with different crystal symmetry. \cite{singh2016PCCP, BiSOC_PRB2007} Moreover, D\'{\i}az-S\'anchez et al.\cite{BiSOC_PRB2007} have reported that the dynamical properties and the interatomic force constants of Bi are very sensitive to the strength of SOC. They reveal that SOC softens the phonon modes in Bi by about 10\% and yields remarkable agreement when compared to that of the experimental values. However, SOC has much smaller effects on the lattice parameters. \cite{BiSOC_PRB2007} In a similar work, Serrano et al. \cite{SbSOC_PRB2008}  have studied the effects of SOC on the specific heat, the lattice parameter, and the cohesive energy of Sb. Their calculations reveal that all these quantities depend almost quadratically on the SOC strength. \cite{SbSOC_PRB2008} The small change in the $C_{ij}$ values due to SOC can be attributed to the above mentioned reasons. The calculated elastic constant values are consistent with an experimental work reported by Lichnowski et al.\cite{Lichnowski1976}, where they investigated the elastic properties of Bi$_{1-x}$Sb$_{x}$  ($0.03 < x < 0.1$) single crystals for small Sb concentration.

\begin{table*}[hbt!]
\centering
\caption{List of elastic constants ($C_{ij}$) calculated with (PBE+SOC) and without SOC (PBE). $C_{ij}$ values (in GPa units) calculated with PBE+SOC are given in parentheses. $x$ represents the concentration of Bi in  Bi$_{x}$Sb$_{1-x}$. The space group of each composition is given in the square brackets. \\ }
\begin{adjustbox}{width=1.0\textwidth, center=\textwidth } 
    \begin{threeparttable}
    \setlength{\arrayrulewidth}{0.3mm}
\setlength{\tabcolsep}{6pt}
\renewcommand{\arraystretch}{1}
\begin{tabular}{lrrrrrrrrrrrrrr}
  \hline
  {\bf Composition} & $x$ & $C_{11}$ & $C_{22}$ & $C_{33}$ & $C_{44}$ & $C_{55}$ & $C_{66}$ & $C_{12}$ & $C_{13}$ & $C_{23}$ & $C_{15}$ & $C_{25}$ &  $C_{35}$ &$C_{46}$  \\
  \hline
\multirow{3}{*}{Sb [166]}  & \multirow{3}{*}{0.0} & 92.2 &  & 35.8 & 29.8 &  &   & 21.8	& 20.3 & & & & & \\
				& 		       & (89.3) &  & (38.9) & (28.3) &  &  & (22.5) & (20.4) & & & & & \\ 
\hspace{1cm} Theory \tnote{a}		& 		       & 91 &  & 38 & 27 &  &  & 24 & 21 & & & & & \\ 

\hline\\
\multirow{2}{*}{Bi$_{1}$Sb$_{7}$ [06]}  & \multirow{2}{*}{0.125} & 98.5 & 83.1 &  36.2 &	36.8 &  13.0 & 16.2  &  6.4  & 12.7  & 32.0 & 15.1 & 6.4 & 1.9 & 6.8 \\
				&							&	(95.1) &	(81.5) &	(37.4) &	(34.0) &	(13.7) &	(15.2) &	(7.2) &	(13.0) &	(30.9) & (13.2) & (5.4) & (2.3) & (6.5) \\ 
\hline\\
\multirow{2}{*}{Bi$_{1}$Sb$_{1}$ [160]}  & \multirow{2}{*}{0.5} & 75.5 &  & 29.1 & 13.4 &  &  & 21.8 & 18.0 & & & &  & \\
				& 		       & (68.7) &  & (31.2) & (12.4) &  &  & (22.6) & (19.6) & & & & & \\
\hline\\
\multirow{2}{*}{Bi$_{3}$Sb$_{1}$ [160]}  & \multirow{2}{*}{0.75} & 67.3 &  & 31.0 & 7.8 &  &  & 27.0 & 20.7 & & & & &  \\
				& 		       & (58.0) &  & (34.0) & (8.3) &  &  & (25.4) & (21.8) & & & & & \\ 
\hline\\
\multirow{2}{*}{Bi$_{7}$Sb$_{1}$ [08]}  & \multirow{2}{*}{0.875} & 61.6 &  63.4 & 26.3 & 4.2 & 5.8 &  20.7  & 23.1 & 16.9 & 17.8 & 1.6 & -4.1 & -0.6 & -4.3 \\
				& 		            & (54.6) &	(58.1) &	(29.5) & (3.9) & (4.5) & (18.3) & (23.2) & (18.7) & (18.6) & (0.63) & (-3.3) & (0.6) & (-4.2) \\
\hline\\
\multirow{2}{*}{Bi$_{9}$Sb$_{1}$ [08]}  & \multirow{2}{*}{0.9} & 25.7 &  64.5 & 62.4 & 20.5 & 8.1 &  6.6  & 16.5 & 16.8 & 20.9 & -0.5 & -5.5 & 5.7 & -4.5  \\
				& 		       & (30.0) &	(66.3) &	(55.2) & (15.8) & (10.7) & (0.5) & (16.7) & (15.4) & (18.8) & (-4.4) & (-5.6) & (2.9) & (-0.9)  \\
\hline\\
\multirow{2}{*}{Bi [166]}  & \multirow{2}{*}{1.0} & 68.6 &  & 31.7 & 6.0 &  &   & 27.8	& 21.3 & & \\
				& 		       & (62.6) &  & (36.1) & (8.8) &  &  & (25.6) & (23.3) & & \\ 
\hspace{1cm} Theory\tnote{a} 	&    & 68 &  & 30 & 10 &  &  & 24 & 19 &  &  \\
\hspace{1cm} Theory\tnote{b} 	&    & 67.7 &  & 40.6 & 8.7 &  &  & 25.0 & 24.3 &  & \\
\hspace{1cm} Exp.\tnote{c} 	&   & 69.3 &  & 40.4 & 13.5 &  &  & 24.5 & 25.4 & &  \\
\hspace{1cm} Exp.\tnote{d} 	&   & 68.7 &  & 40.6 & 12.9 &  &  & 23.7 &  & &  \\ 

\hline
\end{tabular}
    \begin{tablenotes}
         \item[a] Ref.\cite{Jong2015} data from \href{https://materialsproject.org}{materials project database}.
         \item[b] Ref.\cite{ArnaudPRB2016} data from LDA+SOC calculations. 
         \item[c] Ref.\cite{Lichnowski1976} experiment was performed at 4.2 K. For high temperature $C_{ij}$ values for Bi, see ref.\cite{Lichnowski1976} and references therein.
         \item[d] Ref.\cite{Eckstein1960} experiment was performed at 4.2 K. 
     \end{tablenotes}
 \end{threeparttable} 
\end{adjustbox}
\end{table*} 

Notably, the strength of SOC in Sb is much smaller compared to that of in Bi, consequently, changes in the $C_{ij}$ values for Sb-rich compositions are relatively less (slightly over our convergence criteria of 1.0 GPa) compared to that of in the Bi-rich compositions. However, we do notice considerable SOC induced changes in the $C_{ij}$ values for Bi-rich compositions. The influence of SOC on the $C_{ij}$ (to increase or decrease the $C_{ij}$ and therefore the stiffness) is fairly uniform across all compositions. Even pure Sb and Bi follow the same trends with the exception of $C_{12}$. These trends indicate that, in general, due to the SOC effects Bi$_{x}$Sb$_{1-x}$ becomes less stiff along the $x$ and $y$ major axes for deformations along $x$ and $y$ directions, more stiff along the $z$ major axis for deformation along the $z$ direction, and they differ for the transverse forces and responses in the $x-y$ plane. In general, SOC effects cause elastic softening in all directions perpendicular to the $z$ axis. The observed elastic softening could be associated to the SOC induced softening of phonon modes. \cite{BiSOC_PRB2007} In a previous work, Arnaud et al.\cite{ArnaudPRB2016} investigated the effect of SOC on the elastic properties of Bi and observed a similar SOC induced elastic softening. Their reported values are in good agreement with our data presented in Table 1. Here, we would like to note the peculiar effect of SOC on the Bi rich compound, Bi$_{9}$Sb$_{1}$. In Bi$_{9}$Sb$_{1}$ we see that the aforementioned trends are reversed for most of the $C_{ij}$ values. Also, anomalous changes in the elastic stiffness constants can be observed for Bi$_{9}$Sb$_{1}$, which will be discussed in more detail later. These changes support the before mentioned complex relationship between the SOC and the electronic and phonon bandstructure, leading to directional changes in the bonding within the material.

Our calculations indicate that SOC causes small changes (overall less than 1.0 \%) in the Bi-Sb, Bi-Bi, and Sb-Sb bond lengths, which when combined with the phonon softening could be held accountable for the observed SOC induced changes in the $C_{ij}$ values. The maximum variation in the bond length due to SOC is within the range of $\pm$0.03 \AA. Further details of the bond-lengths, lattice parameters, electronic bandstructure, and phonon bandstructure of all studied structures can be found in ref. \cite{singh2016PCCP} Since elastic constants are defined in terms of free energy with respect to strain, the following conclusion can be made here: in presence of SOC, electrons in material are redistributed to minimize the total free energy, thereby recovering some of the strain energy and reducing the effective elastic stiffness.

\begin{figure*}[htb!]
 \centering
 \includegraphics[width=18cm, keepaspectratio=true]{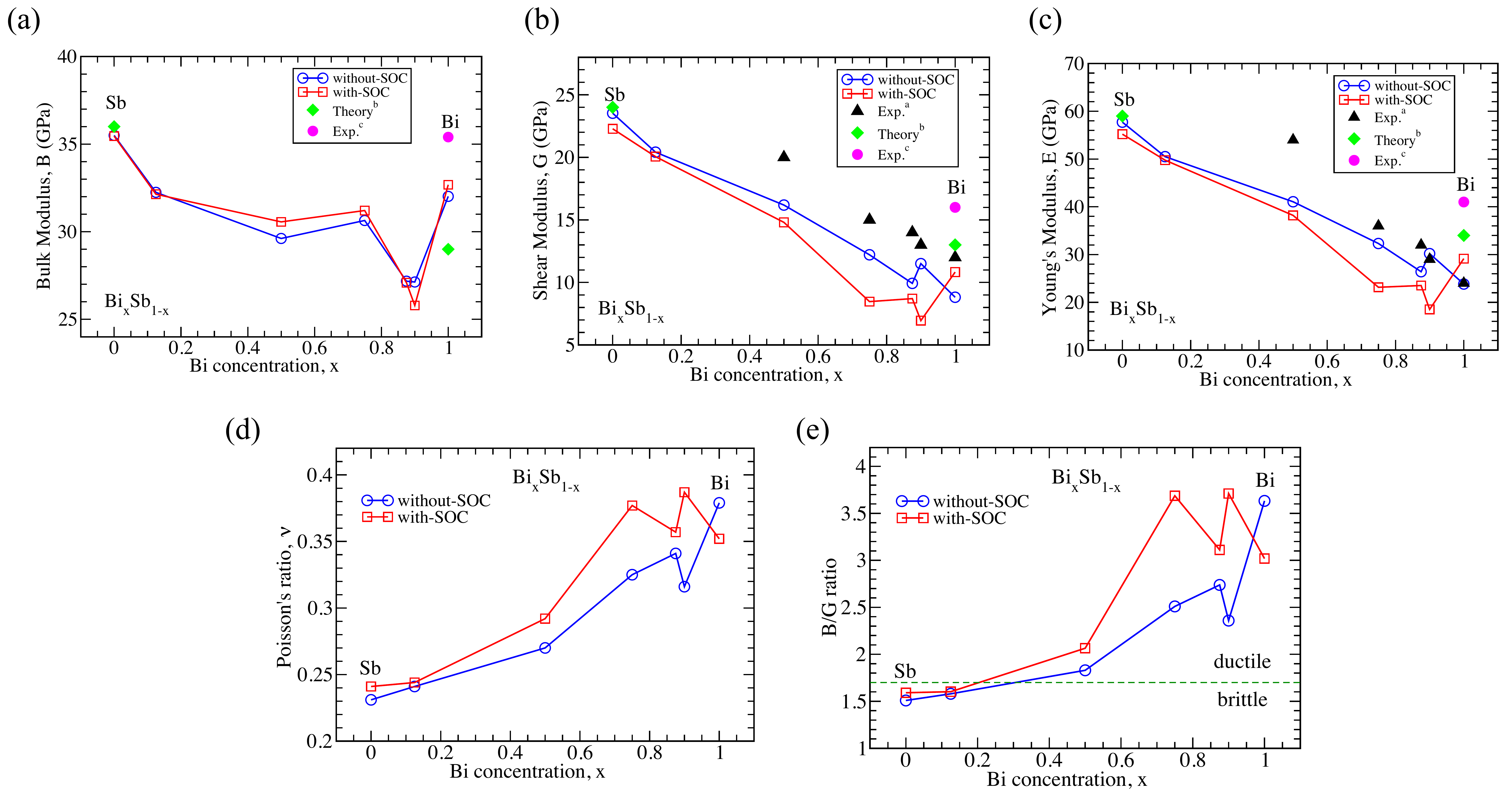}
 \caption{(Color online) Mechanical properties of Bi-Sb binaries calculated with and without inclusion of SOC (a) Bulk modulus $B$ (in GPa), (b) Shear modulus $G$ (in GPA), (c) Young's modulus $E$ (in GPA), (d) Poisson's ratio $\nu$, and (e) $B/G$ ratio. Green dotted line in Fig. (e) shows the boundary ($B/G$ = 1.7) below (above) which material behaves as brittle (ductile). \\ 
 $^a$ Experimental data at room temperature from ref.\cite{Gopinathan1974} \\  
 $^b$ Theoretical data from ref.\cite{Jong2015} \\ 
 $^c$ Experimental data at 4.2 K from ref.\cite{Lichnowski1976} }
\label{fig:EC}
\end{figure*}

 \subsection{Mechanical properties}

We further test the mechanical stability of all the studied structures. A material can be considered mechanically stable if it passes the Born-Huang mechanical stability criteria. \cite{born1955dynamical, nye1985physical} This criteria states that in order to be mechanically stable, the Gibbs free energy of any relaxed crystal, {\it i.e.} in absence of any external load, must be minimum compared to any other state reached by means of an infinitesimal strain. This requires that the elastic stiffness matrix $C_{ij}$ is positive-definite,  {\it i.e.} all the eigenvalues of $C_{ij}$ are positive and the matrix is symmetric. Additionally, all the leading principle minors and any arbitrary set of minors (trailing minors) of $C_{ij}$ must be positive. If a crystal, regardless of its symmetry, satisfies the aforementioned conditions, it can be considered mechanically stable. The mathematical expressions for these conditions have been reported for different crystal classes by various research groups. \cite{PatilPRB2006, ZhijianPRB2007, WU_PRB2007, MouhatPRB2014} It is important to mention here that in some of the published papers \cite{PatilPRB2006, ZhijianPRB2007, WU_PRB2007} these conditions are incorrectly generalized from the cubic criteria (specially for the lower symmetry structures), which could lead to wrong quantitative analysis. However, it could not change the qualitative picture of mechanical stability of a crystal. For the first time, Mouhat and Coudert correctly generalized the Born-Huang mechanical stability conditions for all crystal classes. \cite{MouhatPRB2014} Therefore, we refer the reader to the seminal paper of Mouhat and Coudert for further details regarding the necessary and sufficient conditions for the mechanical stability conditions. \cite{MouhatPRB2014} In our case, we find that all the studied Bi-Sb binary structures pass the Born-Huang mechanical stability test and hence can be considered mechanically stable. 

Once the $C_{ijkl}$ constants are calculated, the four moduli ($B$, $G$, $E$, and $\nu$) can be obtained by using relations between the constants.\cite{Pugh1954} Details of these relations for different crystal systems, and for Voigt and Reuss bound schemes are summarized in ref.\cite{WU_PRB2007}  The bulk modulus represents the volume compressibility of the material and is given by $\rm{B}$ = $\rm{E}/{3(1-2\nu)}$. \cite{nye1985physical} Young's modulus gives a measure of the stiffness of the system. It is simply ratio of the stress along an axis to strain along that axis. A material is very stiff if it has large $E$. Poisson's ratio is used as a measure of plasticity as it measures the expansion of material in the transverse direction to the direction of compression. It is calculated using $\nu = (3B-2G)/2(3B+G)$. The shear modulus, or the modulus of rigidity, describes the deformation of the system under transverse internal forces. It is related to Young's modulus and the Poisson ratio by  $\rm{G}=\rm{E}/2(1+\nu)$.  A way to measure the brittleness or ductility of a material comes from the ratio of the bulk modulus to the shear modulus, $B/G$ ratio, with values above 1.7 giving ductile behavior. \cite{Pugh1954, Pavlic2017} 

Figures~\ref{fig:EC} (a-c) show the observed variation in the $B, G,$ and $E$ values as a function of Bi-concentration in Bi$_{x}$Sb$_{1-x}$. Red (Blue) color represents the data points calculated with (without) inclusion of SOC. We notice that $B, G,$ and $E$ values systematically decrease with increasing Bi-concentration, however, change in $B$ is relatively less compared to that for $G$ and $E$. As expected, the effects of SOC are more dominant towards Bi-rich side than that of towards Sb-rich side. We notice that in all moduli except $B$, these effects on the Bi-rich compositions are present, and the same reversal in trends mentioned in the previous section can be seen in Bi$_{9}$Sb$_{1}$. 

The available experimental and theoretical data (given in Fig.~\ref{fig:EC}) are in excellent agreement with our theoretical calculations.\cite{Gopinathan1974, Jong2015, Lichnowski1976} Here, it important to mention that all the theoretical values are lower than that of the experimental observations. This is due to the fact that we used GGA approximation in all our calculations, and GGA is well-known to underestimate the elastic constant values. \cite{WU_PRB2007} We also notice that the Poisson's ratio ($\nu$) and $B/G$ ratio increase with increasing Bi-concentration, indicating increase in the ductile behavior of Bi-rich compositions. This could be associated to decrease in the strength of the covalent bonds in Bi-rich compositions. The Bi-Bi bond length in pristine Bi (3.10 \AA) is considerably larger compared to the Sb-Sb bond length in pristine Sb (2.96 \AA), thus suggesting stronger covalent bonding in Sb. The average bond length increases with increase in the Bi-concentration. The bond lengths are as follows: in pristine Sb: Sb-Sb bond = 2.96 \AA; in Bi$_{1}$Sb$_{7}$: Sb-Sb bond = 2.98 \AA; in Bi$_{1}$Sb$_{1}$: Bi-Sb bond = 3.04 \AA; in Bi$_{3}$Sb$_{1}$: Bi-Sb = 3.03 \AA~and Bi-Bi = 3.09 \AA; in Bi$_{7}$Sb$_{1}$: Bi-Sb = 3.02 \AA~and Bi-Bi = 3.10--3.12 \AA; in Bi$_{9}$Sb$_{1}$: Bi-Sb = 3.05 \AA~and Bi-Bi = 3.12 \AA, and in pristine Bi (Bi-Bi = 3.10 \AA). Consequently, the increasing bond-length causes decrease in the elastic moduli and increase in the $\nu$ and $B/G$ values. The observed variation in the mechanical properties is consistent with changes in the bond-length. Thus, monotonic decrease in $B, G$, and $E$ values with increasing Bi-concentration can be correlated with decreasing valence electron density. \cite{MCGaoJPCM2013} However, the anomalies observed in the elastic properties of Bi$_{9}$Sb$_{1}$ composition with low-Sb concentration that  are consistent with previous experimental studies\cite{Gopinathan1974} warrant for a further investigation of this issue. The first-principles calculations using Virtual Crystal Approximation (VCA), which are beyond the scope of the present work, can offer good insights to resolve this issue. The possible reasons behind the observed anomalies in the properties of Bi$_{9}$Sb$_{1}$ are discussed below.

 \subsection{Negative Poisson's ratio}
 
\begin{figure*}[htb!]
 \centering
 \includegraphics[width=18cm, keepaspectratio=true]{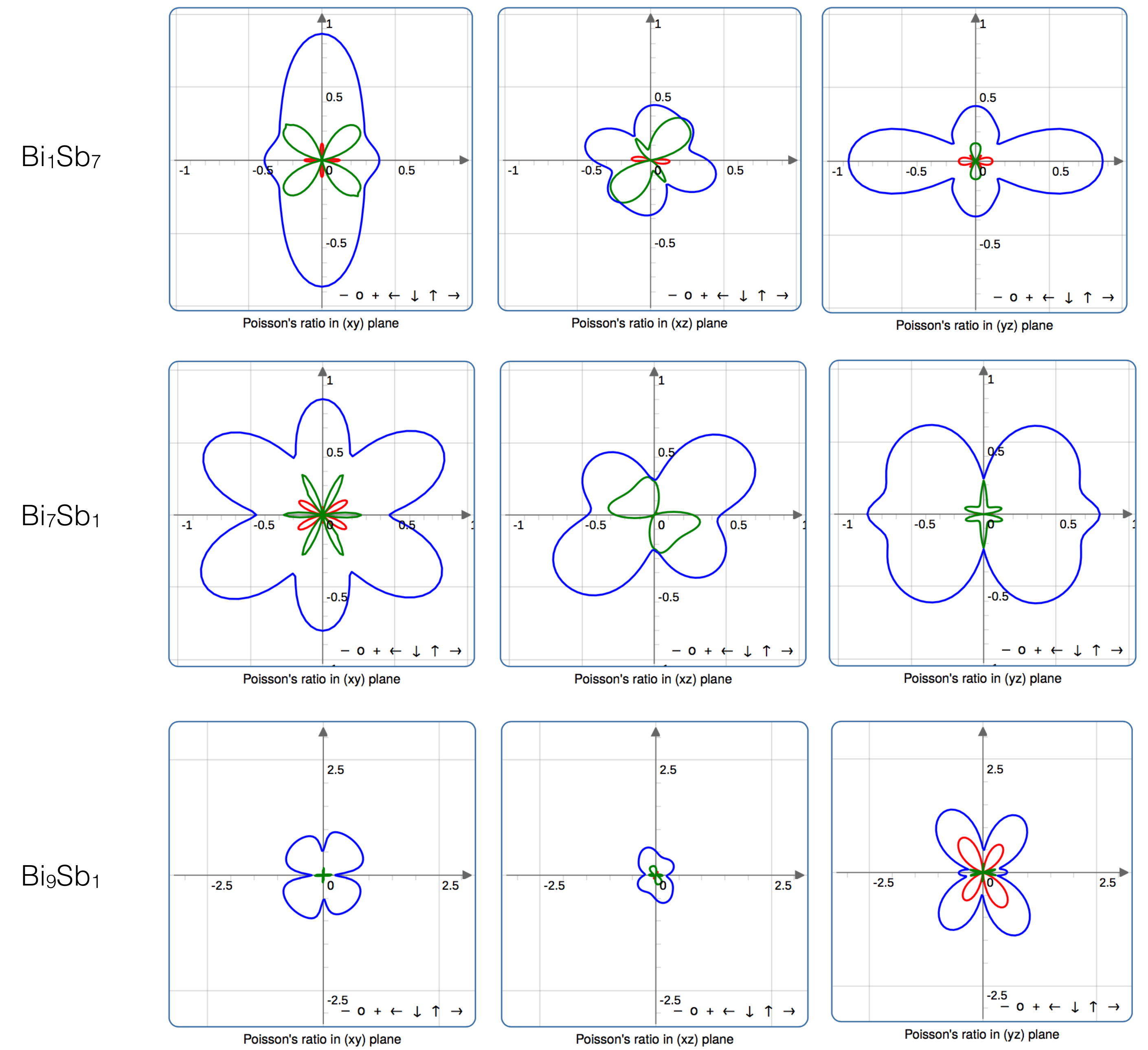}
 \caption{(Color online) Top, middle, and bottom panels represent the calculated Poisson's ratio of Bi$_{1}$Sb$_{7}$, Bi$_{7}$Sb$_{1}$, and Bi$_{9}$Sb$_{1}$ binaries, respectively. All plots were generated using the {\sc ELATE} software\cite{elate}. Green (red) color corresponds to the positive (negative) values of $\nu$ (see text for details). }
 \label{fig:poisson}
\end{figure*}

Materials having negative Poisson's ratio ($\nu$), known as auxetic materials, have attracted special attention of researchers due to their exceptional advantages in the sensing technology.\cite{KittingerPRL1981, LAKES1038, Baughman1998, Baughman2003, Greaves2011} As we mentioned earlier, a positive Poisson's ratio defines the ratio of the transverse contraction to the longitudinal extension of a material during the stretching process. Therefore, materials with negative Poisson's ratio, auxetic materials, are expected to expand in the transverse direction when stretched in the longitudinal direction. Auxetic materials are quite rare in nature as compared to the non-auxetic materials. However, Baughman et al.\cite{Baughman1998} reported that the auxetic property is often observed in cubic elemental metals. Interestingly, auxetic materials with lower symmetry are more appealing for technological applications because they yield much larger strain amplification as compared to the highly symmetric auxetic materials.\cite{Baughman2003} In order to analyze the auxeticity of the studied structures, we thoroughly investigate the elastic tensor of each studied composition calculated with-SOC. 

Using the open source {\sc ELATE} software tool,\cite{GaillacJPCM, elate} we have analyzed the spatial variation of Poisson's ratio for each studied structure. We observe that three out of seven binary structures exhibit significantly large negative Poisson's ratio along different spatial directions. All these structure belong to the low symmetry (monoclinic) space groups, therefore, these structures are more advantageous for technological applications. The results are given in Fig.~\ref{fig:poisson}. Regarding the theoretical details of these plots, we refer the reader to the excellent paper of Gaillac et al.\cite{GaillacJPCM} In spherical coordinates, the determination of $\nu$ requires an extra dimension in addition to the $\theta (0, \pi)$ and $\phi (0, 2\pi)$ coordinates, i.e. $\nu(\theta, \phi, \chi)$ The additional dimension can be characterized by an angle $\chi (0, 2\pi)$.\cite{GaillacJPCM, MARMIER2010} The blue color in Fig.~\ref{fig:poisson} represents the surface obtained at the maximum of $\chi$, whereas the green (red) lobes corresponds to the positive (negative) values of $\nu$ obtained at the minimum of $\chi$.

\subsection{Elastic wave velocities, Debye temperature and Melting temperature}

Knowledge of the elastic wave velocities, Debye temperature and melting temperature is important for practical applications. Therefore, we estimate these quantities using the MechElastic code.\cite{MechElastic} We calculate the longitudinal ($v_l$), transverse ($v_t$), and average ($v_m$) elastic wave velocities using the following relations: \cite{ANDERSON1962, screiber1973elastic}

\begin{equation}
v_{l} = \sqrt{\frac{3B + 4G}{3\rho}},
\end{equation}

\begin{equation}
v_{t} = \sqrt{\frac{G}{\rho}},
\end{equation}

\begin{equation}
\frac{1}{v_{m}} =  \bigg[\frac{1}{3}\bigg(\frac{2}{v_{t}^3} + \frac{1}{v_{l}^3}\bigg)\bigg]^{-1/3},
\end{equation}

where $B$ and $G$ are the bulk and shear moduli, and $\rho$ is the density of material. Conversely, one can also determine the elastic stiffness constants by measuring the distance traveled by an ultrasonic wave pulse and the corresponding time. 

Debye temperature ($\Theta_{D}$) is another important parameters that we can estimate from the knowledge of the elastic wave velocities and the density of material. Debye temperature correlates with several important physical properties such as specific heat, elastic constants, ultrasonic wave velocities and melting temperature. At low temperatures, acoustic phonons are the only vibrational excitations that contribute to the specific heat. Therefore, at low temperatures the Debye temperature calculated from the elastic constants is same as the $\Theta_D$ obtained from the specific heat measurements. We calculate $\Theta_D$ using the following equation: \cite{ANDERSON1962}
\begin{equation}
\Theta_D = {\frac{h}{k_B}\bigg[\frac{3q}{4{\pi}}\frac{N\rho}{M} \bigg]^{1/3}}v_m,
\end{equation}
where $h$ is the Planck's constant, $k_B$ is the Boltzmann's constant, $q$ is the total number of atoms in cell, $N$ is the Avogadro's number, $\rho$ is the density, and $M$ is the molecular weight of the solid. The melting temperature was estimated using the empirical relation: $T_{melt}$ = $607 + 9.3B \pm 555$. \cite{Johnston1996}

\begin{figure}[htb!]
 \centering
 \includegraphics[width=9cm, keepaspectratio=true]{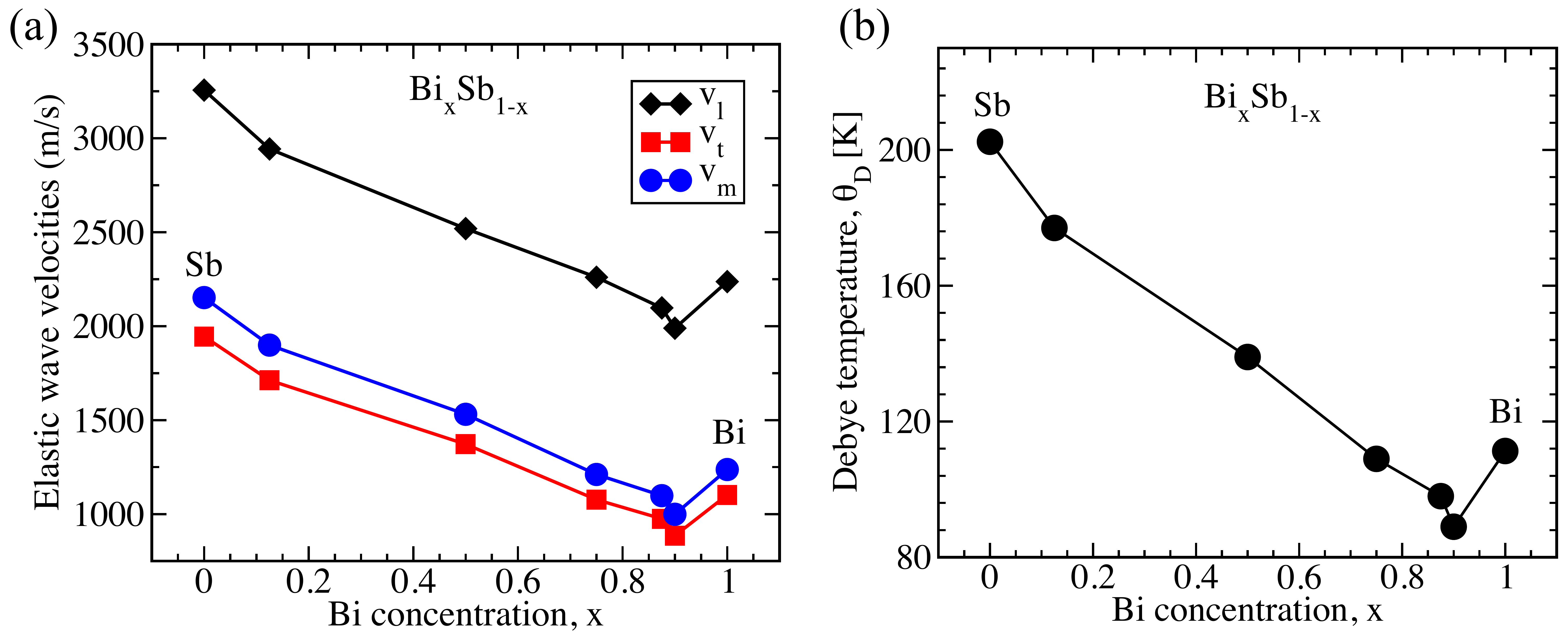}
 \caption{(Color online) (a) Elastic wave velocities, and (b) Debye temperature ($\Theta_D$) of Bi-Sb binaries calculated with SOC.}
 \label{fig:debye}
\end{figure}

Table~2 contains a list of the $v_l$, $v_t$, $v_m$, $\Theta_D$ and $T_{melt}$ values calculated with and without inclusion of SOC. We notice that the magnitude of the elastic wave velocities and $\Theta_D$ decreases due to the SOC effects, which can be associated to the SOC-induced elastic softening. Figure~\ref{fig:debye} shows variation in the elastic wave velocities and $\Theta_D$ as a function of the Bi-concentration. In general, we observe a monotonic decrease in the mentioned quantities with increasing Bi-concentration. However, anomalies from the monotonic trend can be noticed at low Sb-concentration. This observation is consistent with the previous experimental studies of Gopinathan et al.\cite{Gopinathan1974} and  Lichnowski et al.\cite{Lichnowski1976}, where authors investigated the elastic properties of Bi$_{x}$Sb$_{1-x}$ crystals using ultrasonic waves. The experimental $\Theta_D$ values for pristine Sb and pristine Bi at low temperature are $\sim$210~K and $\sim$112~K, which are in excellent agreement with our theoretical findings. 

Anomalous changes in the electronic, thermal, elastic, and mechanical properties of Bi$_{x}$Sb$_{1-x}$ at very low Sb-concentration have been often noted in experiments. Bi$_{x}$Sb$_{1-x}$ undergoes a semimetal-semiconductor phase transition in the Sb-concentration range: $0.07 < x < 0.22 $, and a topological non-trivial insulator phase appears due to the inverted ordering of bands at the L-point of Brillouin zone.\cite{LERNER1968, Brandt1972, Kane_bisb_2008, Hseih2008} Rogacheva et al.\cite{ROGACHEVA2008580} studied the effect of low Sb-concentration on the lattice parameters, microhardness, electrical conductivity, magnetoresistance, and the Seebeck coefficient of Bi$_{x}$Sb$_{1-x}$. Their experiments revealed an anomalous change in the properties of Bi$_{x}$Sb$_{1-x}$ at small $x$ values, which were attributed to the percolation transition, geometric re-ordering of atoms, and semimetal-semiconductor electronic phase transition. They further argued that at low Sb-concentration, the elastic fields of neighboring atoms begin to overlap causing partial compensation of elastic stress with reversed signs, which leads to an abrupt decrease in the elastic stiffness of the entire crystal.\cite{ROGACHEVA2008580} Due to this reason, at low Sb-concentration Bi$_{x}$Sb$_{1-x}$ exhibits rapid decrease in the microhardness, electrical conductivity, and Seebeck coefficient. Increasing Sb-concentration beyond a critical value yields formation of new atomic ordering causing enhancement in the elastic and mechanical properties of Bi$_{x}$Sb$_{1-x}$. Same argument can be used to explain the observed variation in the elastic wave velocities and Debye temperature of Bi$_{x}$Sb$_{1-x}$ with varying $x$ (see Fig.~\ref{fig:debye}).

\begin{table}[hbt!]
\centering
\caption{List of the longitudinal ($v_l$), transverse ($v_t$), average ($v_m$) elastic wave velocities, Debye ($\Theta_{D}$) and melting temperatures (T$_{melt}$) calculated with (PBE+SOC) and without SOC (PBE). Values calculated with PBE+SOC are given in parentheses. The space group of each composition is given in square brackets. \\ }
\begin{adjustbox}{width=0.48\textwidth} 
\begin{threeparttable}
    \setlength{\arrayrulewidth}{0.3mm}
\setlength{\tabcolsep}{6pt}
\renewcommand{\arraystretch}{1}
\begin{tabular}{lrrrrr}
  \hline
  {\bf Composition} & $v_{l}$ (m/s) & $v_{t}$ (m/s) & $v_{m}$ (m/s) & $\Theta_{D}$ (K) & T$_{melt}$ (K) \\
  \hline
\multirow{3}{*}{Sb [166]} & 3240 & 1946 & 2152 & 202.4 & 930  \\
				&      (3256) & (1945) & (2152) & (202.4) & (937) \\
\hspace{1cm} Exp.\tnote{$a$}	&  &  & &  209.6 \\ 
\hspace{1cm} Exp.\tnote{$b$}	&  & & &  210.0 \\ 
\hspace{1cm} Exp.\tnote{$c$}	&  & & &  211.3 \\ 
\hspace{1cm} Exp.\tnote{$d$}	&  & & &  211.3 \\ 

\hline\\
\multirow{2}{*}{Bi$_{1}$Sb$_{7}$ [06]}  & 2941 & 1714 & 1902 & 177.2 & 904 \\
							&    (2943) & (1712) & (1899) & (177.0) & (906) \\
\hline\\
\multirow{2}{*}{Bi$_{1}$Sb$_{1}$ [160]}  & 2548 & 1432 & 1593 & 145.0 & 881 \\
							&    (2519) & (1372) & (1531) & (139.3) & (887) \\
							
\hline\\
\multirow{2}{*}{Bi$_{3}$Sb$_{1}$ [160]}  & 2308 & 1151 & 1291 & 116.5 & 893 \\
							&    (2260) & (1077) & (1211) & (109.2) & (893) \\ 
 
\hline\\
\multirow{2}{*}{Bi$_{7}$Sb$_{1}$ [08]}   & 2123 & 1050 & 1179 & 105.0 & 856 \\
							&    (2097) & (975) & (1098) & (98.0) & (863) \\ 

\hline\\
\multirow{2}{*}{Bi$_{9}$Sb$_{1}$ [08]}   & 2158 & 1131 & 1265 & 112.5 & 850 \\
							&    (1989) & (885) & (999) & (88.9) & (847) \\ 

\hline\\
\multirow{2}{*}{Bi [166]}   & 2197 & 1071 & 1203 & 108.2 & 901 \\
				 &    (2237) & (1102) & (1237) & (111.3) & (908) \\
 \hspace{1cm} Exp.\tnote{$e$}	&  &  & &  112 & \\ 

\hline
\end{tabular}
    \begin{tablenotes}
         \item $^{a}$ Ref.\cite{Blewer1968}, $^{b}$ Ref.\cite{McCollum1967}, $^{c}$ Ref.\cite{Culbert1967} data from specific heat measurements at low temperatures.
         \item $^d$ Ref.\cite{White1972}  data from thermal expansion measurements. 
         \item $^e$ Ref.\cite{Fischer1978} data evaluated from the sound velocity measurements.  
     \end{tablenotes}

\end{threeparttable}
\end{adjustbox}

\end{table} 

\subsection{Specific heat}

After evaluating the elastic properties and mechanical stabilities of the Bi-Sb binaries, we focus our attention on the specific heat ($C$) of crystal lattice. Before we start our discussion, it is important to mention that at low temperatures ($T$) the difference between $C_{p}(T)$ (at constant pressure) and $C_{v}(T)$ (at constant volume) is almost negligible and it lies within the uncertainty range of the experiments.\cite{Cardona2007} Therefore, we do not make any distinction between $C_{p}(T)$ and $C_{v}(T)$ in the present work. In the low $T$ limit, the general relationship between $C(T)$ and $T$ can be described using the following expression: 
\begin{equation}
C(T) = \gamma T + \beta T^{3} + \alpha T^{-2}
\end{equation}
where, first and second terms correspond to the electronic and crystal lattice contributions to the specific heat, whereas, the last term addresses the interaction of the nuclear quadrupole moment with the electric field gradient of electrons and lattice. The last term is very small even at low temperatures, however, it might become substantial below 1~K. \cite{SbSOC_PRB2008} Usually at low $T$, the specific heat follows $T^3$ power law due to the dominant contribution from the lattice vibrations. Therefore, plotting $C(T)/T^3$ versus $T$ is a good way to determine the contribution of the lattice vibrations in the net heat capacity. \cite{KremerGaN2005, SbSOC_PRB2008, Cardona2007, Cardona2009} The peak appearing in this plot is the evidence of the deviation from the Debye behavior that is known to separate the contribution of the acoustic phonons and optical phonons in the total specific heat of material. From the observed position of peak in $C(T)/T^3$ versus $T$ plot, one can estimate the Einstein's oscillator temperature which is typically equal to $\sim$ 6$T_{0}$, where $T_{0}$ is the temperature corresponding to the maximum $C(T)/T^3$.\cite{Cardona2007, Cardona2009}

\begin{figure}[htb!]
 \centering
 \includegraphics[width=8.5cm, keepaspectratio=true]{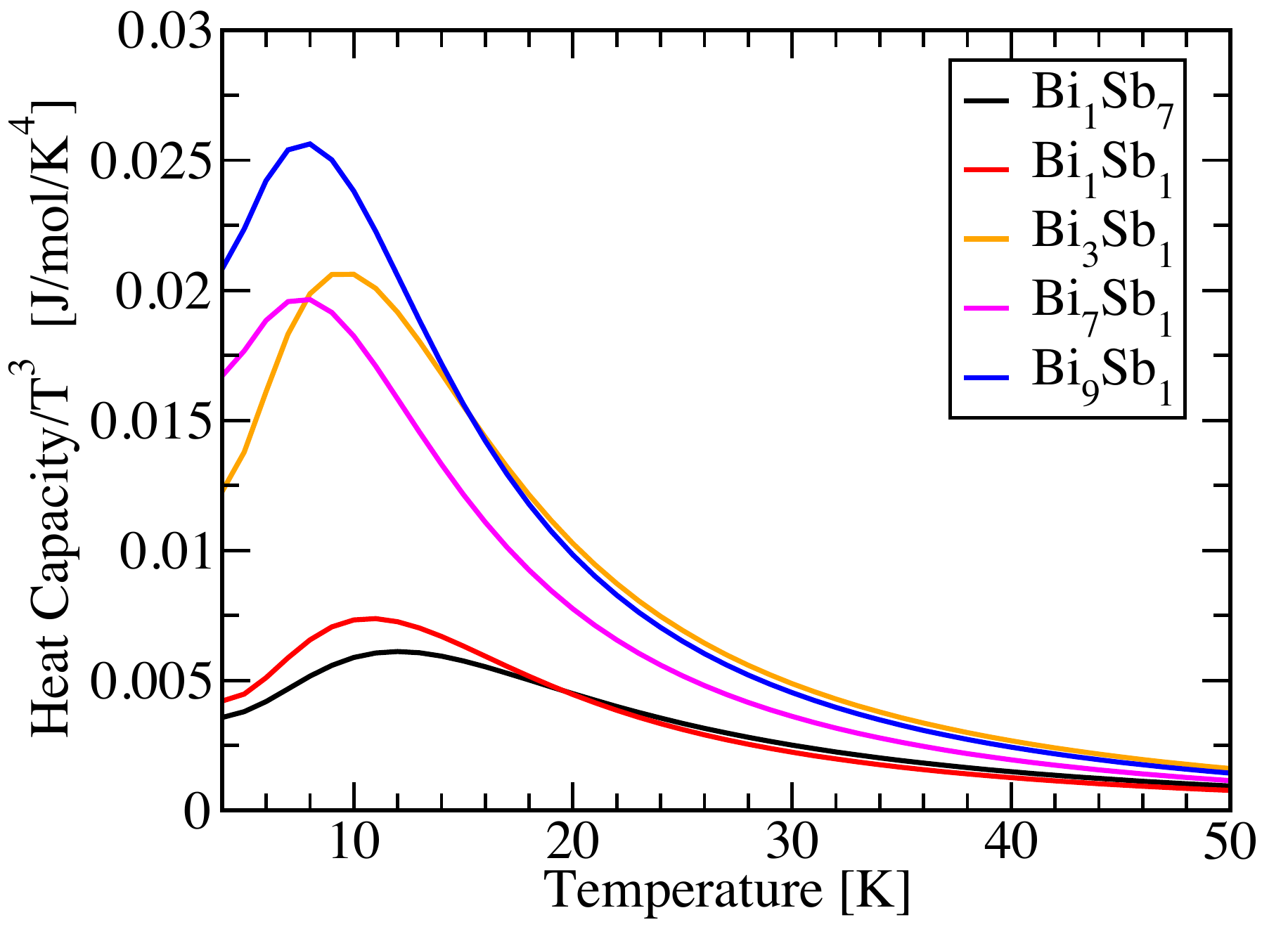}
 \caption{(Color online) $C(T)/T^3$ versus Temperature $T$ data for Bi-Sb binaries. }
  \label{fig:specificheat}
 \end{figure}

Figure~\ref{fig:specificheat} shows the $C(T)/T^3$ versus $T$ plots for all the studied binaries. SOC was not included in the calculation of $C(T)$. The corresponding phonon dispersion for each structure is given in ref.\cite{singh2016PCCP} Noticeably, the peak height in $C(T)/T^3$ versus $T$ plot significantly increases (more than three times) with increasing Bi-concentration from Bi$_{1}$Sb$_{7}$ to Bi$_{9}$Sb$_{1}$. Also, the peak shifts towards lower $T$ with increasing Bi-concentration indicating decrease in $\Theta_{D}$ as we go towards Bi-rich side. This is consistent with $\Theta_{D}$ obtained from the elastic constants calculations, and it can be associated to the decrease in the average strength of the covalent bonds in Bi-rich binaries. We further compare our results with the available theoretical and experimental reports on the pristine Bi\cite{KeesomBi1054, CetasPRB1969, Archer1995, BiSOC_PRL2007}, pristine Sb\cite{DeSorbo1953, SbSOC_PRB2008} and Bi-Sb binaries\cite{Lichnowski1976}. Our findings are in remarkable agreement with the reported data in the literature. Lichnowski and Saunders\cite{Lichnowski1976} have experimentally observed decrease in $\Theta_{D}$ with increasing Bi-concentration. The accepted $T_{0}$ values for pristine Sb and pristine Bi are 7.5~K and 14~K, respectively. \cite{SbSOC_PRB2008, BiSOC_PRL2007} In agreement, we also notice an overall shift of the $T_{0}$ towards lower temperatures from 12~K (for Sb-rich composition) to 8~K (for Bi-rich composition).

Finally, we would like to make a remark about the effect of SOC on the specific heat of the studied binaries. Undoubtedly, SOC is expected to have significant effects on the thermodynamic properties of Bi-rich binaries. In particular, the SOC effects in Bi cause an enhancement in the $C(T)/T^3$ peak height and decrement in the $\Theta_{D}$ value by $\sim$1~K, thereby reducing the discrepancies between the experimental heat capacity data and {\it ab initio} results.\cite{BiSOC_PRL2007} However, SOC is found to have negligible effects on the thermodynamic properties of pristine Sb.\cite{SbSOC_PRB2008} This is the reason why our results (calculated without inclusion of SOC) for Sb-rich binaries compare well with the experimental observations, while there exists a small inconsistency in the data of Bi-rich compositions (for example-- Bi$_{9}$Sb$_{1}$). Including SOC effects in the calculations would yield better agreement with the experimental data, specifically for Bi-rich compositions. Nevertheless, SOC induced changes in the $T_{0}$ values are expected to be within $\pm$1 K range. 

\section{Conclusion}
In this work, we have investigated the elastic, mechanical and thermodynamic properties of several energetically stable Bi-Sb binary structures. We find that bulk, shear, and Young's moduli increase with increasing Sb-concentration in Bi$_{x}$Sb$_{1-x}$, and decrease as we move towards Bi-rich side. However, Poisson's ratio and $B/G$ ratio increase with increasing Bi-concentration suggesting more ductile behavior in Bi-rich compositions. Our calculations reveal that Bi$_{1}$Sb$_{7}$, Bi$_{7}$Sb$_{1}$, and Bi$_{9}$Sb$_{1}$ monoclinic structures exhibit negative Poisson's ratio indicating auxeticity along different spatial directions. We also probe the effect of SOC on the elastic and mechanical properties of Bi-Sb binaries. In general, the SOC effects cause elastic softening in most of the studied structures which can be ascribed to the fact that in presence of SOC, electrons are redistributed to minimize the total free energy, thereby recovering some of the strain energy and reducing the effective elastic stiffness. Our calculations reveal that Debye temperature and magnitude of the elastic wave velocities monotonically decrease with increasing Bi-concentration. This can be ascribed to the decreasing strength of covalent bonds ({\it i.e.} larger bond-length) in the Bi-rich compositions. However, we observe some anomalies in the elastic properties of Bi-rich composition Bi$_{9}$Sb$_{1}$, which requires further investigation. The peak of $C(T)/T^3$ shifts towards lower temperatures and increases in height with increasing Bi-concentration. We find that SOC plays an important role in the determination of the properties for Bi-rich compositions, while the effects of SOC are very small for Sb-rich compositions.  Our overall results are consistent with the available experimental data.

\textit{\\ Acknowledgments}: This work used the Extreme Science and Engineering Discovery Environment (XSEDE), which is supported by National Science Foundation grant number OCI-1053575. Additionally, the authors acknowledge the support from Texas Advances Computer Center (TACC), Bridges supercomputer at Pittsburgh Supercomputer Center, and Super Computing Systems (Spruce and Mountaineer) at West Virginia University. A. H. R. and S. S. acknowledge the support from National Science Foundation (NSF) DMREF-NSF 1434897 and DOE DE-SC0016176 projects. S. S. also acknowledges the support of the Jefimenko Fellowship and Robert T. Bruhn Research Award at West Virginia University.

\bibliography{./bibliography}
\end{document}